\documentclass[%
 reprint,superscriptaddress,
 amsmath,amssymb,
 aps,twocolumn,pra
]{revtex4-1}

\usepackage{graphicx}
\usepackage{multirow}
\usepackage{amsfonts}
\usepackage{verbatim}
\usepackage[width=.8\textwidth]{caption}

\usepackage[ pdftex, plainpages = false, pdfpagelabels,
                 pdfpagelayout = useoutlines,
                 bookmarks,
                 bookmarksopen = true,
                 bookmarksnumbered = true,
                 breaklinks = true,
                 linktocpage=all,
                 pagebackref=false,
                 colorlinks = true,
                 linkcolor = BrickRed,
                 urlcolor  = blue,
                 citecolor = BrickRed,
                 anchorcolor = green,
                 hyperindex = true,
                 hyperfigures
                 ]{hyperref}
\usepackage[usenames, dvipsnames]{xcolor}
\usepackage{xifthen}

\newcommand{\eqn}[1]{\begin{eqnarray} #1 \end{eqnarray}}
\newcommand{\tit}[1]{\textit{#1}}
\newcommand{\tbf}[1]{\textbf{#1}}
\newcommand{\trm}[1]{\textrm{#1}}
\newcommand{\tr}[1]{  \textrm{tr}\left[ #1 \right]  }

\newcommand{\zum}[2]{\displaystyle\sum_{#1}^{#2}}

\newcommand{\onestage}{{One}}
\newcommand{\twostage}{{Two}}

\newcommand{\arxiv}[2][]{\ifthenelse{\isempty{#1}}{\href{http://arxiv.org/abs/#2}{{\tt arXiv:\allowbreak{}#2}}} {\href{http://arxiv.org/abs/#2}{{\tt arXiv:\allowbreak{}#2 [#1]}}}}

\begin{document}

\title{Born's rule as a quantum extension of Bayesian coherence}

\author{John B. DeBrota\smallskip}
\affiliation{\href{https://sites.google.com/view/tuftsqi}{QIQC Group, Department of Physics and Astronomy}, Tufts University, 574 Boston Avenue, Medford MA 02155 }
\affiliation{
\href{http://www.physics.umb.edu/Research/QBism/}{QBism Group, University of Massachusetts Boston}, 100 Morrissey Boulevard, Boston MA 02125, USA}
\affiliation{Stellenbosch Institute for Advanced Study (STIAS), Wallenberg Research Center at Stellenbosch University, Marais Street, Stellenbosch 7600, South Africa}
\author{Christopher A. Fuchs\smallskip}
\affiliation{
\href{http://www.physics.umb.edu/Research/QBism/}{QBism Group, University of Massachusetts Boston}, 100 Morrissey Boulevard, Boston MA 02125, USA}
\affiliation{Stellenbosch Institute for Advanced Study (STIAS), Wallenberg Research Center at Stellenbosch University, Marais Street, Stellenbosch 7600, South Africa}
\author{Jacques L. Pienaar\smallskip}
\affiliation{
\href{http://www.physics.umb.edu/Research/QBism/}{QBism Group, University of Massachusetts Boston}, 100 Morrissey Boulevard, Boston MA 02125, USA}
\affiliation{Stellenbosch Institute for Advanced Study (STIAS), Wallenberg Research Center at Stellenbosch University, Marais Street, Stellenbosch 7600, South Africa}
\author{Blake C. Stacey\smallskip}
\affiliation{
\href{http://www.physics.umb.edu/Research/QBism/}{QBism Group, University of Massachusetts Boston}, 100 Morrissey Boulevard, Boston MA 02125, USA}





\begin{abstract}
The subjective Bayesian interpretation of probability asserts that the rules of the probability calculus follow from the normative principle of Dutch-book coherence: A decision-making agent should not assign probabilities such that a series of monetary transactions based on those probabilities would lead them to expect a sure loss. Similarly, the subjective Bayesian interpretation of quantum mechanics (QBism) asserts that the Born rule is a normative rule in analogy to Dutch-book coherence, but with the addition of one or more empirically based assumptions---i.e., the ``only a little more'' that connects quantum theory to the particular characteristics of the physical world. Here we make this link explicit for a conjectured representation of the Born rule which holds true if symmetric informationally complete POVMs (or SICs) exist for every finite dimensional Hilbert space.  We prove that an agent who thinks they are gambling on the outcomes of measurements on a sufficiently quantum-like system, but refuses to use this form of the Born rule when placing their bets is vulnerable to a Dutch book.  The key property for being sufficiently quantum-like is that the system admits a symmetric reference measurement, but that this measurement is not sampling any hidden variables.
\end{abstract}

\maketitle

\section{Introduction}

The \tit{Born rule} is a centerpiece of quantum mechanics. The way the Born rule is often described in textbooks is as follows:\ We presuppose a density operator $\rho$ to describe a quantum system and a positive-operator-valued measure (POVM) $\{ D_j\}$ with outcomes $j \in \{1,\dots,J \}$ to describe a measurement on the system. The probability $q(j)$ for outcome $j$ is then given by,
\eqn{ \label{eqn:Born}
q(j) = \tr{\rho D_j} \, .
}
But how do we know which operator $\rho$ and which POVM $\{ D_j \}$ to use in a given experiment? A commonplace view is that once the system and its method of preparation have been specified, there is in principle a uniquely correct choice of $\rho$ that provides the best possible description of the real state of the system. Similarly, it is thought that a unique POVM $\{ D_j \}$ exists in principle, which correctly describes the measuring apparatus.

Despite being the common attitude, this interpretation does not stand up to serious scrutiny. For an investigation into the meaning of the symbols $\rho$ and $\{ D_j \}$ leads us into the long-standing measurement problem, which in turn leads to competing interpretations of quantum theory. Most debate focuses on the interpretation of the quantum state $\rho$, and asks whether the quantum state completely describes reality or represents only a partial description of reality.

The radical possibility that the particular quantum-state assignment has nothing to do with an agent-independent reality is the core idea of the quantum interpretation known as \tit{QBism}. (See Ref.~\cite{Fuchs10a} for the first relatively complete statement of QBism, Refs.\ \cite{FuchsStacey2018} and \cite{FMS} for comprehensive reviews of it, and Ref.~\cite{Stacey2019} for a listing of its divergences from the early ``Quantum Bayesianism'' of Caves, Fuchs, and Schack~\cite{Caves02a}.) The central tenet of QBism is that the quantum state $\rho$, taken by itself, says \tit{nothing at all} about external reality. Instead, quantum states encode probabilities that represent an agent's (the physicist's) subjective degrees of belief about the outcomes of possible future measurements on the system.  Evidence supporting this position includes the mathematical fact that quantum states and measurements don't need to be represented by operators in Hilbert space:\ They can be written directly as sets of probabilities.

To understand the Born rule from this point of view, we begin by suspending our usual tendency to interpret the symbols $\rho$ and $\{ D_j \}$ as  descriptors of a system and the measuring apparatus, and instead see them as they are most directly presented to us: as mathematical symbols, written in ink on a page or in pixels on a laptop screen, that we use for some purpose. This shift in viewpoint entails that we do not immediately leap to some conclusion about what it is that the symbols mean---rather, we must slowly and cautiously approach their real meaning by adopting a new attitude towards them, in which their mathematical \tit{form} is not to be assumed but must be derived by a careful consideration of the symbols' \tit{purpose}. With this in mind, we refocus our attention on what these symbols are used for by the physicist.

In the case of the Born rule, we imagine a physicist who is in possession of a system, which we may think of as a physical object located somewhere in the laboratory, and a measuring device, which we may think of as a box into which a system can be placed and which responds by displaying the measurement outcome $j$ on a visible dial. The physicist wants to be able to predict the value of $j$ as well as possible, in order to attain some practical end, such as testing a hypothesis or building a quantum computer. Since the likelihood of the outcome $j$ is therefore important to the physicist, the symbols $\rho$ and $\{ D_j \}$ acquire their meaning because the physicist uses them to decide which probabilities $q(j)$ are best to assign. The sum total of the physics community's past experiments on systems of this kind, as distilled and captured in the formal rules of quantum theory, indicates that the physicist \tit{should} assign $q(j)$ according to equation \eqref{eqn:Born}.

But why? What if the physicist---whether in a fit of rebellion or due to absent-mindedness---assigns probabilities according to some other rule?  QBism asserts that the physicist should expect to suffer for this transgression, because it represents an \tit{inconsistency} between their beliefs. To wit: On the one hand, the physicist believes they are dealing with a quantum system, as being within the purview of quantum theory, and at the same time they believe the likelihood of the outcomes $j$ is calculated by some method other than that in Eq.\ \eqref{eqn:Born}. Both sets of beliefs cannot simultaneously be justified (this is a theorem that we shall prove), so something has to give---but what?

As with many things in life, when tension becomes unbearable, there is no telling which part of the structure will be the first to fail. We can identify at least three possibilities:
\begin{itemize}
\item[(i)]
The physicist may realize that they are not justified in believing the system to be within the purview of quantum theory;
\item[(ii)]
The physicist may decide that their expectations about the outcome $j$ are not justified in light of experience, and may revise $\rho$ or $\{ D_j \}$ or both;
\item[(iii)]
Worst of all, the physicist may carry on with their task oblivious to the inconsistency until---after much labor wasted in failing to achieve their goal---they re-examine their assumptions along the lines of (i) or (ii).
\end{itemize}

This example illustrates that it would be of great use to have a means of detecting inconsistencies in one's beliefs without having to waste the effort of testing them by costly practical experiments. Fortunately there is such a method.

The first step is to ``unpack'' the physicist's beliefs in terms of their probability assignments to the outcomes of hypothetically possible experiments. These thought experiments must be plausible but need not actually be carried out. The next step is to interpret these probabilities as quantities of money that the agent would be prepared to wager on the values of the corresponding experimental outcomes.

\tit{Remark:} ``Money'' is being used here as an abstraction or a metaphor for \tit{any} kind of utility the agent might seek to attain, and whose loss would be undesirable to the agent~\cite{BernardoSmith1994}. The essential point is that the subjective interpretation must supply a reason for using the symbols of probability theory. While the prospect of facing a sure loss of utility sounds a little abstract, we trust that most of our readers can understand why an agent would want to avoid losing money, and therefore why they would want to make their bets in accordance with the probability calculus.

After this unpacking of beliefs into probabilities and then into wagers, the next step is to check for the existence of a \tit{Dutch book}: a series of wagers, each justifiable on the basis of some belief, but whose totality amounts to a certain loss of money regardless of which outcomes actually occur~\cite{deFinetti1990, MISAK}. If the agent finds that such a Dutch book can be made against them, they may conclude that their beliefs are mutually inconsistent, and can proceed to revise them.

This assertion---that a Dutch book implies inconsis\-tency---is called the principle of \tit{Dutch-book coherence}~\cite{deFinetti1990}. It depends on the idea that an agent would not \tit{want} to lose money. That is, it connects the abstract idea of ``inconsistency'' with the concrete and meaningful consequence of ``losing money''.

The principle of Dutch-book coherence is a powerful tool in the subjective Bayesian approach to probability theory. Among other things, it can be used to derive the standard rules of the probability calculus from first principles. Constraints on an agent's probability assignments derived from Dutch-book coherence are called \tit{normative} constraints, to emphasize that no law of nature forces an agent to adhere to them. So it is with the rules of the probability calculus: No law forces us to obey them, but we ignore them at our own risk.

We shall use the principle here to prove that a decision-making agent (like the physicist in our example) who believes a system to be ``quantum'' must then assign probabilities $q(j)$ in accordance with the Born rule \eqref{eqn:Born} through some choice of $\rho$ and $\{ D_j \}$, or else be vulnerable to a Dutch book. This then establishes the Born rule itself as a normative rule, which an agent should use in addition to the rules of the standard probability calculus whenever they are dealing with quantum systems.

To establish our thesis, we must accomplish four things. First, we must unpack the meanings of the symbols $\rho$ and $\{ D_j \}$ in terms of probabilities that the agent assigns to hypothetical experiments, and use the resulting expressions to express the Born rule purely as a constraint on the agent's probability assignments; this will be covered in Sec.\ \ref{sec:norm}\@.
Second, we must unpack the agent's belief that ``the system is quantum'' in terms of the agent's probability assignments to the hypothetical experiments. Evidently we cannot take this to imply that the agent uses the full-blown structure of quantum theory or the Born rule, for this would commit the error of assuming what we set out to prove. Instead we must make use of some minimal assumptions about what ``quantumness'' might mean for the agent's probability assignments. Our particular choice of assumptions is discussed in Sec.\ \ref{sec:core}\@.
Third, we must show how these minimal assumptions, plus Dutch-book coherence, implies the Born rule. This is a straightforward but nontrivial mathematical theorem that we prove in Sec.\ \ref{sec:born}\@. In principle we could stop there, for if we hold fixed the agent's belief that ``the system is quantum'' (as represented by our minimal assumptions), then to not use the Born rule would necessarily mean a transgression of Dutch-book coherence. For the sake of completeness, we will show how to explicitly construct a Dutch book in Sec.\ \ref{sec:dutch}.

\section{The Born rule as a normative constraint on probability assignments \label{sec:norm}}

In this section we review a standard result from the QBist literature, showing how the Born rule can be interpreted as a constraint on the agent's probability assignments~\cite{Fuchs13a,Fuchs15b}. This approach to the Born rule involves a generalization of the double-slit experiment where the ``which way'' measurement is \tit{informationally complete}~\cite{DeBrota20a,DeBrota21}. In the familiar treatment of quantum interference, one compares probabilities calculated for a scenario where there is only one detector (say, an electron counter at a given position) to the probabilities calculated for a scenario that also includes an intermediate measurement (like a device that indicates which slit the electron passed through). Designate the final detector by $\mathcal{D}$ and the optional intermediate detector by $\mathcal{S}$. When an agent Alice sets out to study the phenomenon of interference, she contemplates two alternative experiments:\ Either she sends a system directly to $\mathcal{D}$ and obtains an outcome $j$ (Experiment \onestage), or she passes the system through $\mathcal{S}$ and then $\mathcal{D}$, obtaining some outcomes $i,j$ in succession (Experiment \twostage).

First, consider Experiment \twostage. Let $p(i)$ be the agent's probability to obtain $i$ in the $\mathcal{S}$ measurement and let $R(j|i)$ be her probability to obtain $j$ in the $\mathcal{D}$ measurement conditional on obtaining $i$ in the preceding $\mathcal{S}$ measurement. Elementary probability theory implies that her joint probability to obtain result $i$ followed by $j$ must be equal to $R(j|i)\, p(i)$, and also that the probability to obtain $j$, ignoring the value of $i$, must be:
\eqn{
s(j) = \zum{i}{}\, R(j|i)p(i) \,
}
This relation is commonly known as the Law of Total Probability.

Next consider Alice's probabilities $q(j)$ to obtain $j$ in Experiment \onestage. In what manner are these related to, or constrained by, the probabilities $p(i)$ and $R(j|i)$ that she has already assigned to Experiment \twostage?

It is not immediately obvious that the two sets of assignments should have anything to do with each other; after all, we are talking about different experiments. Nevertheless, there is an assumption of physical similarity between the two, namely that they differ only in the inclusion or exclusion of the $\mathcal{S}$ measurement, and so our expectations about one might well be connected with our expectations of the other. Our task is to make this connection explicit, and show how it depends upon what amount to be \tit{physical assumptions}.  (Despite a common trope of the philosophers and science journalists, there is only so much subjectivity in QBism.)

To begin with, suppose Alice believes that the passage of the system through apparatus $\mathcal{S}$ should not affect her thinking about the system in any way. This can be rephrased as the belief that, for the purposes of assigning probabilities to the outcome of $j$ alone, Alice considers it irrelevant whether that outcome was obtained as part of Experiment \twostage{} or Experiment \onestage, i.e.\ the outcome of the $\mathcal{D}$ measurement is insensitive to whether or not the system was previously sent through apparatus $\mathcal{S}$ or not. Evidently this implies
\eqn{
q(j) &=& s(j) \, \nonumber \\
&=& \zum{i}{}\, R(j|i)p(i) \, .
}
That is, the likelihood of $j$ in Experiment \onestage{} must equal the likelihood of $j$ in Experiment \twostage, which was given by the Law of Total Probability above.

Note that the validity of this relation depends on a substantive assumption about how the apparatus $\mathcal{S}$ affects, or does not affect, the system. Different beliefs about whether $\mathcal{S}$ changes the whole experimental context may imply a rule of a different form. It is instructive to ask what is the \tit{most general} form such a rule could take.

To answer this question, it will be useful to use a vector space representation of the probability assignments. Let $p$ be an $N$-vector with components $\{ p(i) : i=1,\dots,N \}$ and $R$ be an $J\times N$ matrix with components $\{ R(j|i) \}$.  We will further make some assumptions about the operational setting of these experiments. Al\-though we shall present them as background assumptions without much fanfare, the reader should be alert to their importance for everything to follow.

First, we assume that the agent assigns $p$ and $R$ independently of one another. To see why, consider a fixed choice of apparatus $\mathcal{S}$. The probabilities $p$ she assigns to the system going into this apparatus might reasonably only depend upon her thinking about the system itself, and not on what subsequent measurement $\mathcal{D}$ she might choose to do in the future. Conversely, we assume that the device $\mathcal{S}$ can be modeled as a measure-and-reprepare device~\cite{DeBrota20b}, that is, the system emerging from it only depends on the outcome $i$ that was produced, and not on any further details of the system that entered it. Since the probabilities in $R$ are all conditioned upon the outcome $i$, they ought to depend only on the choice of device $\mathcal{D}$, and not on the beliefs about the outcomes if the system were instead passed into $\mathcal{S}$. In conclusion, $p$ and $R$ can be treated as independently chosen expressions of beliefs.

Secondly, we assume that the measurement outcomes $j$ are \tit{noncontextual}. That is, for the purposes of assigning the probability $q(j')$ to an individual outcome $j=j'$, only the row of $R$ having components $\{R(j'|i):i=1,\dots,N \}$ should be relevant. To justify this, consider two measuring apparatuses $\mathcal{D}_1$ and $\mathcal{D}_2$ which share the same outcome $j'$. What can it mean to say that this outcome is ``the same,'' given that it appears on the dials of two different devices? A priori, the outcomes should be given different labels, say $j'_1$ and $j'_2$. However, if Alice assigns these outcomes the same probabilities given \tit{any} any other beliefs she might hold of the system, then she is justified in identifying them as equivalent, and can represent them using a single label $j' := j'_1 = j'_2$. In other words, in assigning $q(j')$ to a single label $j'$ that can appear on the dials of different apparatuses, it is implicit that the agent must consider it irrelevant which of the apparatuses the outcome $j'$ belongs to. If it had been relevant, then she would not have seen fit to assign them the same label~\cite{Fuchs02a}. For brevity, we will use the notation $r_j$ to stand for the column vector whose transpose is the $j$th row of $R$.

We shall now augment the innocuous premises that we have made so far with our first substantial assumption:\\

{\bf A1}. \tit{Existence of an Informationally Complete Apparatus}. There exists a choice of apparatus $\mathcal{S}$ for which the agent's assignments to Experiments \onestage{} and \twostage{} are related by a rule of the form:
\eqn{ \label{eqn:nosigma}
q(j) = \mathcal{F}( p, r_j) \, .
}
That is, all relevant differences between the two experiments are fully captured by the agent's beliefs $p(i)$ and $R(j|i)$ about the system and the apparatuses involved, without the need for any additional parameters $\sigma$.\\

{\bf A1} makes a nontrivial statement about the nature of the physical world, since it asserts that a certain kind of measurement is physically possible in principle. To formalize this idea, let us fix a choice of apparatus $\mathcal{S}$ satisfying {\bf A1}, which we call the \tit{reference apparatus}. A state is a vector of probabilities $p$ for the possible outcomes that may be generated by applying this apparatus to the system. (If it has not already been apparent, this distinguishes the idea of a \tit{state} from that of the \tit{system}, by which we mean the physical object the agent takes an action upon by way of the apparatus.)

For each outcome $i$ of the reference apparatus $\mathcal{S}$, we can consider a ``double-pass'' through it, i.e., taking the measured system and passing it again through $\mathcal{S}$ to obtain another outcome $k \in \{1,\dots,N \}$. In our notation, this will generate a conditional probability $R(k|i)$, where $\mathcal{S}$ itself now plays the role of $\mathcal{D}$.  Supposing an initial uniform distribution for the $i$'s, we can use Bayes' rule to formally invert to a conditional probability $p(i|k)$.  We shall call these probabilities the \tit{reference states} for the reference apparatus and give them the special notation $\{e_k(i) : k = 1,\dots,N \}$.

Now that the assumptions of our general probabilistic setting have been carefully laid out, we can contemplate what the Born rule would imply in this setting. Let us suppose for the moment that our agent is cognizant of quantum theory and makes the following associations for his system and apparatuses:  a density operator $\rho$ for the system, a completely general POVM $\{D_j\}$ for the apparatus $\mathcal{D}$, an informationally complete POVM $\{E_i\}$ consisting of rank-1 elements for the apparatus $\mathcal{S}$, which finally in turn gives a L\"uders rule collapse to one of the pure states $\{\Pi_i\}$ upon its execution. Quantum theory dictates that the probabilities the agent should assign to $j$ in Experiment \onestage{} are
\eqn{ \label{eqn:otherBorn}
q(j) = \tr{\rho D_j} \, ,
}
which is just the Born rule. Similarly, the outcome probabilities for the informationally complete POVM $\{E_j\}$ are
\eqn{ \label{eqn:SIC-rep-p}
p(i) = \tr{\rho E_i} \, ,
}
and the conditional probabilities for the second stage of Experiment \twostage{} are
\eqn{ \label{eqn:SIC-rep-r}
R(j|i) = \tr{\Pi_i D_j} \, .
}
So far so good, but this does not answer the question as we have posed it. Recall that on the Bayesian account the agent \tit{begins} with the probability assignments $p(i)$ and $R(j|i)$ in Experiment \twostage, and is asked to deduce $q(j)$ from these. If we could somehow invert Eqs.\ \eqref{eqn:SIC-rep-p} and \eqref{eqn:SIC-rep-r} to obtain $\rho$ and $\{ D_j\}$ in terms of the $p(i)$ and $R(j|i)$, we could substitute those expressions into \eqref{eqn:otherBorn} and have the solution.

To see how this is done, simply note that $\{E_i : i=1,\dots,N \}$ has the defining property that its elements span the space of linear operators on the system's Hilbert space. In dimension $d$, this requires $\mathcal{S}$ to have a minimum of $d^2$ outcomes, and if $N=d^2$ it is then called a \tit{minimal} IC-POVM, or \tit{MIC}~\cite{DeBrota20c}.

In discussions of QBism it has been customary to assume the existence of an apparatus whose representation is not only minimal but is also \tit{symmetric}, meaning the elements are proportional to rank-1 projectors and the overlaps between distinct elements are constant:
\eqn{
\tr{E_i E_j} = \frac{d \, \delta_{ij} + 1}{d^2(d+1)} \, \quad  \forall i, j \, .
}
This is called a \tit{SIC-POVM}, or simply a \tit{SIC} (pronounced ``seek'')~\cite{ZAUNER_PHD,RENES04}. In our probabilistic setting, the symmetry property amounts to the requirement that the reference states satisfy
\eqn{ \label{eqn:prob_symmetry}
e_j(i) = (1-Nc) \delta_{ij} + c \, \quad  \forall i, j \, ,
}
for some constant $c$.

Although it is not known whether SICs exist in all dimensions, exact algebraic constructions have been found in over 100 dimensions and high-precision numerical solutions have been found in nearly 100 more~\cite{Grassl20}.  Beyond this raw evidence, there are also a number of rather elegant mathematical reasons to suggest that SICs \tit{ought} to exist in all finite dimensions \cite{Appleby11,Appleby15,Appleby17,Bengtsson20,DeBrota20d,Pandey20}.  In fact, it is widely believed that it is only a matter of time before a full existence proof will be found.  (For a broad review this topic, see Ref.\ \cite{SIC_REVIEW}.) In light of this, QBism usually takes it for granted that the reference apparatus may be supposed to implement a SIC, especially as this would give the Born rule a very special and simple mathematical expression in terms of probabilities~\cite{DeBrota20b,Fuchs2017}, which we are about to see.

Proceeding with the convention that $\mathcal{S}$ is associated with a SIC, the formulas \eqref{eqn:SIC-rep-p} and \eqref{eqn:SIC-rep-r} can be inverted to obtain:
\eqn{ \label{eqn:SIC-rep-rho}
\rho = \zum{i=1}{d^2}\, \left((d+1)p(i)-\frac{1}{d} \right) \Pi_i \, ,
}
where $\{ \Pi_i \} := \{ d E_i\}$, and
\eqn{ \label{eqn:SIC-rep-D}
D_j = \zum{i=1}{d^2}\, R(j|i)\left((d+1)E_i-\frac{1}{d}\, \mathbb{I} \right) \, .
}
with $\mathbb{I}$ the identity operator. Substituting Eqs.\ \eqref{eqn:SIC-rep-rho} and \eqref{eqn:SIC-rep-D} into \eqref{eqn:otherBorn} yields the QBist version of the Born rule,
\eqn{ \label{eqn:Urgleichung}
q(j) = \zum{i=1}{d^2}\, \left( (d+1)p(i)-\frac{1}{d} \right) R(j|i) \, ,
}
which we can see amounts to a special choice of the function $\mathcal{F}$ appearing in {\bf A1}. Hereafter, we shall simply call the expression in Eq.\ \eqref{eqn:Urgleichung} ``the Born rule,'' despite the fact that it has not been completely established in quantum theory, for lack of a conclusive proof of SIC existence.

This provides the Born rule with an operational meaning: It is an example of a rule that an agent might use to relate their probability assignments between hypothetical Experiments \onestage{} and \twostage. Why would an agent use this particular rule instead of some other? The answer is that they would use it if they believed the system and the apparatuses possess some of the essential features we normally use full-blown quantum theory for. In the next section we unpack the meaning of this statement in terms of the agent's probability assignments.

\section{Minimal assumptions for quantum systems \label{sec:core}}

Given a reference apparatus $\mathcal{S}$ of $N$ outcomes, the physically legitimate states are represented by the set of assignments $p(i)$ that the agent considers to be possible; we denote this set $\mathcal{P}^N$. Similarly, we let $\mathcal{R}^J$ denote the space of physically possible measurement apparatuses with $J$ outcomes. Specifically, the elements of $\mathcal{R}^J$ are sets of conditional probabilities $\{R(j|i) : \forall j,i \}$ the agent considers to be plausible probabilities for the outcomes $j$ conditional on sending the $i$-th reference state $e_i(k)$ into the apparatus $\mathcal{D}$. We shall assume that, among the physically possible apparatuses, there is a ``garbage disposal'' apparatus whose outcome is uniformly distributed regardless of its input, i.e.\ to which the agent assigns the uniform distribution $\{ R(j|i)=\frac{1}{J} :  \forall j,i \}$.

The basic rules of probability theory, which follow from Dutch-book coherence, say that a probability vector $p$ must have nonnegative entries and be properly normalized. Moreover, the entries in a matrix of conditional probabilities must be nonnegative, and each column of the matrix must sum to 1. As mentioned above, noncontextuality lets us split up a matrix $R$ that defines a measurement and consider its rows separately. We will write the $j$-th row of a measurement matrix $R$ as the transpose of a column vector $r_j$. The bare minimum requirement imposed by Dutch-book coherence alone on such vectors is that their entries lie in the unit interval. We will refer to vectors that meet these basic requirements for probabilities and conditional probabilities respectively as \tit{Dutch-book valid}. When we augment the abstract basic rules of probability theory with lessons about the character of the physical world, we find that only a subset of the Dutch-book valid vectors are \tit{physically valid}. The set $\mathcal{P}^N$ of physically valid states is a proper subset of the probability simplex, and the set $\mathcal{R}^J$ of physically valid $J$-outcome measurements is a proper subset of the set of all $J \times N$ stochastic matrices. We will use $\mathcal{M}^N$ to denote the set of all physically valid vectors $r_j$ that can be used as ``building blocks'' for matrices in $\mathcal{R}^J$.

Let us focus our attention on the set of state assignments $\mathcal{P}^N$ that the agent considers physically valid. Since we consider the probabilities $p(i)$ as components of an $N$-vector, we can define the Euclidean inner product between any two probability $N$-vectors $p_1$ and $p_2$.  We shall assume this inner product to have potentially nontrivial lower and upper bounds:
\eqn{ \label{eqn:ineq}
L \leq (p_1, \, p_2) \leq U \, , \quad \forall p_1,p_2 \in \mathcal{P}^N \, .
}

It will turn out that two states separated by $L$ can always be perfectly distinguished by some measurement \tit{other than} the reference measurement. This is very unlike what happens in a classical probabilistic theory, where all measurements are coarse-grainings of the information one would have if one knew the values of the intrinsic physical degrees of freedom:  Coarse-grainings cannot make probability distributions more distinguishable! A strictly positive value of $L$ is thus a signal of nonclassicality.

The following two assumptions are intended to apply generally, not just to quantum systems. To begin with, we assume that the agent is as permissive as possible about what can be a physical state, within the constraints represented by the inequalities \eqref{eqn:ineq}:\\

{\bf A2}. \tit{Maximality with Respect to the Inequalities}. If a vector $p_1$ satisfies the inequalities \eqref{eqn:ineq} for all $p_2 \in \mathcal{P}^N$, then $p_1 $ also belongs to $\mathcal{P}^N$.\\

Intuitively, {\bf A2} asserts that ``anything not forbidden is permitted'': Any probability assignment that is not already ruled out by the bounds \eqref{eqn:ineq} must in fact be a physically valid state. It is a straightforward consequence of {\bf A2} that $\mathcal{P}^N$ is convex and closed and contains the uniform distribution $\{\frac{1}{N} : i=1,\dots,N\}$.

Because the uniform distribution is a physically valid state, it is intuitively plausible that vectors close to the uniform distribution should be so as well. Heuristically, a theory that is as permissive as possible should include the largest possible region around the uniform distribution within the set of physically valid states $\mathcal{P}^N$. Our next assumption formalizes this by considering a ball of physically valid states surrounding the uniform distribution inside $\mathcal{P}^N$---the so-called ``in-ball''. We assume that the radius of this ball is fixed not by an arbitrary parameter fed into the theory, but by the geometrical constraints of probability theory itself:\\

\noindent {\bf A3}. \tit{Maximality of the In-Ball}. The state space $\mathcal{P}^N$ has an in-ball of the maximum possible size, namely, the size of the largest ball that can be inscribed within the probability simplex.\\

In Appendix \ref{app:vitality2} we show that {\bf A2} implies that $\mathcal{P}^N$ contains a set of $N$ states having the form
\eqn{ \label{eqn:basis_states}
p_k(i) = (1-NL)\delta_{ik} + L \, , \qquad k \in \{1,\dots, N\} \, ,
}
which span $\mathcal{P}^N$. In the same Appendix we also show that {\bf A3} implies that these states have the maximum possible norm,
\eqn{ \label{eqn:extremality}
(p_k,p_k) = U \, , \qquad \forall k \, .
}
Since these have the form \eqref{eqn:prob_symmetry} required of the reference states, we might as well adopt the convention that these \tit{are} the reference states, i.e.\ that when a system is passed into $\mathcal{S}$ and produces the outcome $i$, the agent assigns the state $e_k(i) := p_k(i)$ to the system coming out of $\mathcal{S}$. This is a convenient choice for the reference states because {\bf A2} guarantees that any agent operating within our assumptions must consider these to be physically valid states.

We next introduce another natural property of the state space---the \tit{dimension} $d$ of the system, which depends on both the possible states $\mathcal{P}^N$ and the possible apparatuses $\mathcal{R}^J$. To get at this notion, let us first define what it means for a set of states $\{p_i\}$ to be \tit{mutually maximally distant} (MMD).  We say a set is MMD when for all $p_i$ and $p_j$ in it,
\eqn{
(p_i, p_i) &=& U, \nonumber \\
(p_i, p_j) &=& L,  \quad  i \neq j.
}
The number $d$ is then defined operationally as the maximum size possible for an MMD set.

\tit{Remark:} One can find multiple notions of the dimension of a system in the literature. For example, one could define $d$ as the \tit{measurement dimension}---the maximum number of states that are mutually perfectly distinguishable by a single apparatus \cite{BRU14}. In the present context, our reference measurement is informationally complete, but generally not optimal for distinguishing states in a single shot. Therefore we find it more natural to use a definition of $d$ that refers to the bounds $U$ and $L$ on states' overlap, rather than one that refers to \tit{perfect} distinguishability. Our definition coincides with the measurement dimension in both classical and quantum mechanics, but may not coincide with it in general.

Note that an apparatus with fewer than $d$ outcomes cannot possibly be informationally complete, so in general $N \geq d$. In the special case where $\mathcal{P}$ is a simplex, which we can loosely identify with classical theories, the $N$ vertices of the simplex form a MMD set of maximum size, and it follows that $N=d$ and $L=0$. This suggests that an agent who thinks themselves to be dealing with a nonclassical theory ought to assign $N > d$ and $L > 0$.

More specifically, we can now ask what values of these numbers might represent an agent's belief that they are operating in the quantum regime. Quantum theory tells us that the minimum number of outcomes of any informationally complete apparatus must be equal to $d^2$. An agent who believes that the system and apparatuses are ``quantum'' must therefore at least believe in the following operational constraint: \\

{\bf A4}. \tit{Quantum Prerequisite}. The minimal number of outcomes for the reference apparatus is $N=d^2$.\\

It is worth pointing out that {\bf A4} only refers to quantities with an operational meaning, and so it is entirely plausible that an agent could come to believe {\bf A4} without first knowing, or having to derive, quantum theory.

As one might expect, {\bf A4} has some highly nontrivial consequences for the structure of $\mathcal{P}^N$. For one thing, it implies that no two physically valid states can be perfectly distinguished by the reference measurement, and hence that $L>0$. In fact, in Appendix \ref{app:vitality4} we show that {\bf A2}--{\bf A4} imply $L = 1/(d^2+d)$ and $U = 2L$. Putting these results together implies that the reference states are:
\eqn{ \label{eqn:e_components}
e_k(i) = \frac{1}{(d+1)}\delta_{ik} + \frac{1}{d(d+1)} \, .
}

In what follows we will show that so long as the agent assigns probabilities in accordance with {\bf A1}--{\bf A4}, then the function $\mathcal{F}$ which they choose to mediate their assignments to Experiments \onestage{} and \twostage{} must have the form of the Born rule as expressed in \eqref{eqn:Urgleichung}.

\tit{Remark:} Assumptions {\bf A1}--{\bf A4} appear to be insufficient to imply the full structure of quantum theory, i.e.\ to imply that the state space $\mathcal{P}^N$ is necessarily equivalent to quantum state space. The full quantum state space can be achieved using additional assumptions, as discussed in detail elsewhere \cite{QPLEX}, but since our present aim is only to derive the Born rule, we have no need of those assumptions here.

\section{Operational derivation of the Born rule \label{sec:born}}

We have already seen that {\bf A1} asserts a function $\mathcal{F}$ that constrains the agent's probability assignments to Experiment \onestage{} in terms of the probabilities they would assign to a hypothetical Experiment \twostage. In this section we show that the additional assumptions {\bf A2}--{\bf A4} imply that the function $\mathcal{F}$ has the form of the Born rule in the form Eq.\ \eqref{eqn:Urgleichung}.

The first stage in the proof is to show that the function $\mathcal{F}$ acts linearly on the $r_j$ argument. First we note that $\mathcal{F}$ defines a map
\eqn{
\mathcal{F}: \mathcal{P}^N \times \mathcal{M}^N \mapsto [0,1] \, ,
}
such that for any measurement $\{r_j : j=1,\dots,J \}$ we have:
\eqn{ \label{eqn:frame_condition}
\zum{j}{J} \, \mathcal{F}(p,r_j) = 1 \qquad \forall p \in \mathcal{P}^N \, .
}
We then have the following result:\\

\tbf{Proposition.} Consider any $r' \in \mathcal{M}^N$ that decomposes as a linear combination of vectors in $\mathcal{M}^N$,
\eqn{
r' = \zum{x}{} \, \alpha_k r_x \, \qquad r',r_x \in \mathcal{M}^N \, ,
}
where $\alpha_k$ are real (not necessarily positive) coefficients. Then $\mathcal{F}$ preserves linear combinations on $r$, i.e.\
\eqn{ \label{eqn:linear_r}
\mathcal{F}\left(  p, \zum{x}{} \, \alpha_x \, r_{x} \right) &=& \zum{x}{} \, \alpha_x \, \mathcal{F}\left( p, r_{x} \right) \, ,
}
for any fixed $p \in \mathcal{P}^N$.\\

To prove this, we employ the following mathematical preliminary:\\

\tbf{Lemma.} Consider a function $f: \mathcal{V}^N \mapsto [0,1]$ acting on a spanning subset of vectors $\mathcal{V}^N$ in the space $\mathbb{R}^{N,+}$ of vectors having positive or null components. Furthermore, suppose that $f$ is additive on $\mathcal{V}^N$, i.e.\ given any $v_1,v_2 \in \mathcal{V}^N$ such that $(v_1+v_2) \in \mathcal{V}^N$, we have
\eqn{
f(v_1+v_2) = f(v_1) + f(v_2) \, .
}
It then follows that $f$ preserves linear combinations on $\mathcal{V}^N$, i.e.\
\eqn{ \label{eqn:prop2}
f\left( \zum{x}{} \, \alpha_x v_x \right) =  \zum{x}{} \, \alpha_x \, f\left(v_x \right) \,
}
for real coefficients $\alpha_x \in \mathbb{R}$.\\

\tit{Proof:} See Appendix \ref{app:prop2}.\\

In order to apply this Lemma to our present situation, let us fix $p$ and define $f(r_j) := \mathcal{F}(p,r_j)$. Evidently $f$ is a map from $\mathcal{M}^N$ to the unit interval, and $\mathcal{M}^N$ is a spanning subset of vectors in $\mathbb{R}^{N,+}$ since it contains the vectors defining the reference measurement, Eq.\ \eqref{eqn:e_components}. To be able to apply the Lemma, it remains to show that $f$ is additive.

Suppose that $\{ r_1, r_2, r_3 \}$ is a set of three physically valid vectors that satisfy the normalization constraint and so qualify as a measurement. From this set we can obtain another measurement by coarse-graining the outcomes $j=2$ and $j=3$ into a single outcome $j=4$, resulting in the set $\{r_1, r_4 \}$. Elementary probability theory states that the probability of the union of two mutually exclusive events must be the sum of the probabilities of the individual events, hence
\eqn{
R(j=4|i) = R(j=2|i) + R(j=3|i) \quad \forall i \, ,
}
or in vector form, $r_4 = r_2 + r_3$. Since the $f(r_j)$ must sum to 1, we have
\eqn{
f(r_1) + f(r_2) + f(r_3) &=& f(r_1) + f(r_4) \nonumber \\
\Rightarrow f(r_2) + f(r_3) &=& f(r_2 + r_3) \, ,
}
and so additivity of $f$ is proven. We can therefore apply the Lemma to conclude that for any fixed $p \in \mathcal{P}^N$ we have
\eqn{
\mathcal{F}\left( p, \zum{x}{} \, \alpha_x r_x \right) =  \zum{x}{} \, \alpha_x \, \mathcal{F} \left(p, r_x \right) \,
}
for real coefficients $\alpha_x \in \mathbb{R}$. $\Box$\\

We have argued from noncontextuality that for fixed $p$, the output probability $q(j)$ must be a linear function of $r_j$. This means that we can write $q(j)$ as the inner product of $r_j$ with some vector. This vector depends somehow on $p$, so let us call it $g(p)$. What can we say about the vector-valued function $g$? We know that if the $r_j$ are the vectors that define the reference measurement itself, then $q = p$. So, the inner product of $r_j$ with $g(p)$ just reads out an element of $p$. Whatever $g$ does, it can be undone by a linear transformation! Therefore, $g$ itself is a linear transformation, defined by the inverse of the conditional-probability matrix for the reference measurement.

We can express this in matrix form as
\eqn{ \label{eqn:matrix_urg}
\mathcal{F}(p,r_j) = r_j^T \, \Phi \, p \, ,
}
where $\Phi$ is an $N \times N$ matrix whose inverse is the matrix whose columns are the vectors $e_k$. Recalling that the components of $e_k$ are given by Eq.\ \eqref{eqn:e_components} due to assumptions {\bf A2}--{\bf A4}, we have
\eqn{
\Phi_{ij} = (d+1)\delta_{ij}-\frac{1}{d} \, .
}
Consequently,
\eqn{
\mathcal{F}\left( p(i),R(j|i) \right) &=& \zum{i=1}{d^2}\, \left( (d+1)p(i)-\frac{1}{d} \right) R(j|i) \, , \nonumber\\
}
which is precisely Eq.\  \eqref{eqn:Urgleichung}, i.e.\ the Born rule in probabilistic form, as promised.

The Born rule in Eq.\ \eqref{eqn:matrix_urg} would reduce to the Law of Total Probability if the $\Phi$ matrix were replaced with the identity. The fact that $\Phi$ does not equal the identity is an expression of how quantum theory deviates from classical probability. Prior work has shown that there is no choice of reference measurement which can make a statement of the Born rule that comes closer to the Law of Total Probability~\cite{DeBrota20b}. Deducing Eq.\ \eqref{eqn:matrix_urg} with its specific choice of $\Phi$, as we have done here, derives the \tit{irreducible margin} of nonclassicality exhibited by quantum theory.

\section{A Dutch book for the Born rule \label{sec:dutch}}

In order to understand the Born rule as an addition to Dutch-book coherence, we need to understand how much Dutch-book coherence itself implies. The essential point can be illustrated by reviewing the Dutch-book argument for the additivity of probabilities.

Dutch book arguments like the one we are about to make are an introspective tool that any agent can use to test the consistency of their own probability assignments. To do so, the agent imagines a hypothetical ``bookie'' with whom they negotiate to buy or sell lottery tickets. We can see the additivity requirement arise in the following way.

Alice contemplates two mutually exclusive events, $E$ and $F$. The bookie, who wants to profit off any inconsistencies among her beliefs, offers to sell her lottery tickets whose values are contingent upon these events, and to buy such tickets from her, in arbitrary combinations. The simplest tickets to write are
\eqn{
  T_E := [ \trm{Worth \$1 if $E$} ] \, \nonumber \\
  T_F := [ \trm{Worth \$1 if $F$} ] \, ;
}
and then there is a ticket of slightly more complicated form,
\eqn{
  T_{E \lor F} := [ \trm{Worth \$1 if $E \lor F$} ] \, .
}
Alice declares that she is willing to price these tickets at \$$p(E)$, \$$p(F)$ and \$$p(E \lor F)$ respectively. The bookie then considers the prices Alice has set. If $p(E \lor F) > p(E) + p(F)$, the bookie sees that Alice will agree to the following series of transactions:\\
Buy $T_{E \lor F}$ for \$$p(E \lor F)$;\\
Sell $T_E$ for \$$p(E)$;\\
Sell $T_F$ for \$$p(F)$.\\
Alice, having committed to these prices, pays the bookie \$$p(E \lor F)$ to buy the first ticket and then sells the bookie the other two tickets, so she runs at a loss of \$$(p(E) + p(F) - p(E \lor F))$. If $E$ occurs, Alice wins \$1 for $T_{E \lor F}$, but must pay the bookie \$1 (since the bookie holds \$$T_E$), leaving her still with a net loss. Likewise, if $F$ happens, Alice earns a dollar and loses a dollar, her balance remaining negative. Finally, if neither $E$ nor $F$ occurs, none of the three tickets are worth anything, and Alice's balance again stays at its initial negative value.

On the other hand, if Alice declares her prices and the bookie sees that $p(E \lor F) < p(E) + p(F)$, then the bookie simply exchanges ``buy'' and ``sell'' in the above set of transactions, again forcing Alice into a loss. In brief, no matter what the circumstances, holding the ticket $T_{E\lor F}$ is always equivalent to holding the pair of tickets $T_E$ and $T_F$. Therefore, in order to avoid being Dutch-booked, Alice must gamble in accord with the condition $p(E \lor F) = p(E) + p(F)$.

Dutch-book arguments can also be made that ticket prices should never be negative (Alice knows she would be a fool to pay the bookie to take a ticket off her hands), and that they should be bounded above by 1 (Alice knows better than to buy a ticket for more than it could ever be worth). In brief, the basic rules of probability theory emerge from the requirement that Alice gamble \tit{coherently}, that is, in such a way as to avoid a sure loss. Probabilities simply \tit{are} the gambler's internally self-consistent prices for tickets.

The relation between joint and conditional probabilities
\eqn{
  p(E \land F) = p(E)p(F|E)
  \label{eqn:joint-conditional}
}
is often presented as an axiom, but the Dutch-book method can derive it as a theorem, because the bookie can offer \tit{conditional} lottery tickets that pay off if both $E$ and $F$ occur, but are refunded if $E$ does not:
\eqn{
  T_{F|E} := [ \trm{Worth \$1 if $E \land F$, but refund if $\neg E$} ] \, .
}
If Alice does not relate joint and conditional probabilities in accord with Eq.\ \eqref{eqn:joint-conditional}, then she can be Dutch-booked, because holding the conditional ticket $T_{F|E}$ is equivalent to holding the pair of tickets
\eqn{
  T_{E \land F} := [\trm{Worth \$1 if $E \land F$}] \, , \nonumber\\
  T_{X} := [\trm{Worth \$$p(F|E)$ if $\neg E$}] \, .
}
The events $E \land F$ and $\neg E$ are mutually exclusive, so the additivity rule applies, and thus in order to be Dutch-book coherent, Alice must set
\eqn{
  p(F|E) = p(E \land F) + p(F|E)p(\neg E) \, .
}
This yields Eq.\ \eqref{eqn:joint-conditional} once we recognize that $p(E) + p(\neg E) = 1$, which also easily follows from Dutch-book coherence.

If Alice discovers that she is vulnerable to a Dutch book, say by declaring
\eqn{
  p(E \lor F) > p(E) + p(F) \, ,
}
then she can restore coherence by adjusting any or all of the probabilities $p(E)$, $p(F)$ and $p(E \lor F)$ to establish balance. The mathematics does not say which to modify; that is up to Alice's best judgment.

Living up to the standard of Dutch-book coherence means that in Experiment \onestage, Alice's probabilities for the different outcomes $j$ must divide up the unit interval among them. Likewise, in Experiment \twostage{}, Alice's $p(i)$ must be nonnegative numbers that sum to 1, and her joint probabilities for $i$ followed by $j$ must obey the Law of Total Probability. But Dutch-book coherence alone cannot bridge between Experiment \onestage{} and Experiment \twostage, because the lottery tickets pertaining to the reference measurement are simply inoperative if the reference measurement is not physically performed. To make any connection between Experiments \onestage{} and \twostage, we need at least a little physics!

In the previous section, we identified the necessary physics as the assumptions {\bf A1}--{\bf A4}. When added to the basic rules of probability theory, these conditions pinpoint the Born rule.

Suppose that Alice's gambling commitments for Experiment \onestage{} are internally self-consistent, and so are those she makes for Experiment \twostage, but when put together, they turn out to violate the Born rule. That is, Alice declares vectors $p$ and $q$ along with a matrix $R$ that satisfy the requirements of nonnegativity and normalization, but
\eqn{
  q(j) \neq \zum{i=1}{d^2} \left[(d+1) p(i) - \frac{1}{d}\right] R(j|i) \, .
  \label{eqn:not-born-rule}
}
The message of the Born rule is that Alice should work to remove this inconsistency. However, the quantum formalism does not provide guidance on how exactly to do so. Alice might decide that the reference measurement $\mathcal{S}$ is so central to her thinking that she ought to maintain her expectations about it, namely the vector $p$, and reset $q$ accordingly. On the other hand, she might say that she has much more experience with measurements like $\mathcal{D}$---they could be cheap while the reference measurement $\mathcal{S}$ is expensive---and so it is best to keep $q$ and adjust $p$ and $R$. In other words, the quantum formalism does not help Alice decide, although more ``meaty'' quantum physics could.

Does an inconsistency like Eq.\ \eqref{eqn:not-born-rule} manifest in the possibility of Alice being Dutch-booked? One way to see how it can is to revisit the theme that Alice is vulnerable to a Dutch book if she declares unequal prices for two equivalent sets of tickets.

The bookie, who is quite clever and will go to any length to be adversarial, offers Alice the possibility to gamble upon \tit{her own future declarations of belief}. This is a standard move when constructing Dutch-book arguments for how probabilities might best be \tit{updated} over time \cite{VANF, FS_reflection}. The subject of coherent probability-updating strategies is a level beyond what we have discussed so far, and it offers more flexibility than is often acknowledged \cite{FS_reflection}. Without developing the subject in depth, we can still make good use of the basic gambling-on-probabilities idea. Suppose that Alice thinks over Experiment \twostage{} and then declares her gambling commitments in the form of $p$ and $R$. The bookie asks her if she has accepted the Born rule, and she says that she has. Quickly, the bookie calculates the probability vector $q$ using the Born rule and offers to buy a ticket
\eqn{
  T_q := [ \trm{Worth \$1 if Alice declares $q$} ] \, .
}
Alice, hesitant, fixes her price for $T_q$ at less than \$1. She then works through the calculation and finds that $q$ is the unique probability vector consistent with the Born rule and her declared $p$ and $R$. Chagrined, she declares $q$ and pays the bookie \$1, leaving herself with a net loss.

The lesson illustrated by this scenario is that if Alice accepts the assumptions {\bf A1}--{\bf A4}, then holding tickets about her declaring $p$ and $R$ is equivalent to holding a ticket about her declaring $q$. Assigning unequal prices to equivalent sets of tickets makes her vulnerable to a Dutch book; she can restore coherence by adjusting any of her probability assignments, though probability theory itself does not say which.

\section{Conclusions}

In this paper, we have established that the Born rule in the form Eq.\ \eqref{eqn:Urgleichung} can be viewed as a normative constraint on an agent's probability assignments. It is a normative constraint above and beyond the standard rules of probability theory. On their own, the rules of probability theory do not tell an agent how their probabilities for one experiment (Experiment \twostage) should constrain their assignments to another slightly different experiment in which one of the measurements is missing (Experiment \onestage). To make this connection requires some extra empirically motivated assumptions about the \tit{physics} relevant to these two experiments. We identified a set of such assumptions, the first three of which ({\bf A1}--{\bf A3}) represent general assertions about physical systems and are compatible with both classical and quantum systems, while the last ({\bf A4}) represents a minimal requirement for believing the systems to be essentially quantum in spirit if not letter. We then showed that any agent who adheres to {\bf A1}--{\bf A4} and strives to uphold the principle of Dutch-book coherence must use the Born rule as the constraint that connects their probability assignments between the hypothetical Experiments \onestage{} and \twostage.

We suspect that our set of assumptions can be streamlined---that is, that it will be possible to enumerate fewer assumptions, potentially at the cost of lengthier chains of deductions between them. The most physically significant of the assumptions we have made are {\bf A1}, which entails that probabilities are noncontextual in a sense inherited from Gleason~\cite{BUSCH, RENES}, and {\bf A4}, which pushes the general mathematics in the direction of quantum theory specifically.

The assumption of maximality, {\bf A2}, expresses the ethos that ``everything not forbidden must be allowed''; if a probability vector $p$ were consistent with the upper and lower inner-product bounds \eqref{eqn:ineq} and yet excluded from the theory's state space, then the theory would tacitly be assuming some other physical principle, and we wish to be as parsimonious with our physical principles as we can. Assumption {\bf A3} can also be viewed as an appeal to parsimony, for it amounts to saying that no new distance scale within the probability simplex has to be introduced by hand in order to determine the outer boundaries of the state space.

Much of this work is based on an earlier informal presentation by one of us \cite{VITALITY}, in which it was suggested that the bilinear form of the Born rule could be derived from van Fraassen's  \emph{reflection principle} \cite{VANF, FS_reflection}. Such an approach could potentially circumvent our rather abstract and lengthy proof of linearity, replacing it with a more direct conceptual argument using Dutch-book coherence. This requires a careful study of how the reflection principle applies to \emph{conditional} probabilities, which will be the subject of a subsequent paper~\cite{FutureWork}.

Finally we reiterate that although we have recovered the Born rule within a general setting, we have not gotten all the way from abstract probability theory to quantum mechanics. As we remarked after assumption {\bf A4}, at least one additional condition is required to ensure that a \emph{qplex,} a maximal consistent set with respect to the bounds
\eqn{
\frac{1}{d(d+1)} \leq (p_1,p_2) \leq \frac{2}{d(d+1)}, \nonumber
}
is a \emph{Hilbert qplex} isomorphic to quantum state space. We know that postulating a particular type of symmetry is sufficient, and we have elsewhere conjectured that this condition can be relaxed to a more qualitative one \cite{QPLEX}. In fact, it is possible that a condition with the same ethos as {\bf A2}, asking that the state space have as few ``distinguishing marks and scars'' as mathematically possible, could not only serve this role but also allow a relaxation of {\bf A4} to an even weaker condition like $N > d$ \cite{VITALITY}.

\acknowledgments
This work was supported in part by the John E. Fetzer
Memorial Trust. CAF and BCS were supported by the John Templeton
Foundation. The opinions expressed in this publication are those of
the authors and do not necessarily reflect the views of the John
Templeton Foundation. We thank Darran McManus for comments.


\begin{appendix}
\section{Consequences of maximality \label{app:vitality2}}
In this Appendix we prove the two claims made in Sec.\ \ref{sec:core} about the consequences of assumptions {\bf A2} and {\bf A3} for the state space $\mathcal{P}^{N}$. First we prove that {\bf A2} implies that the state space contains a set of states of the form \eqref{eqn:basis_states}.

Any set of probability vectors that is maximal (according to assumption {\bf A2}) has a property called \tit{self-polarity}, defined as follows. Let $H$ be the hyperplane in $\mathbb{R}^{N}$ consisting of vectors whose elements sum to unity, i.e., the hyperplane of probabilities and quasi-probabilities. The \tit{polar} of a point in $H$ is the set of all points in $H$ whose inner product with the given point is greater than the lower bound $L$ in the inequalities \eqref{eqn:ineq}. The polar of a set of points is the set of all points which are in the polars of all the given points. (This terminology is adapted from the study of polytopes.) It follows from the maximality of $\mathcal{P}^N$ that the polar of any subset of $\mathcal{P}^N$ is also a subset of $\mathcal{P}^N$, that is, $\mathcal{P}^N$ is a \tit{self-polar} set.
Note that the operation of taking the polar reverses inclusion, i.e.\ if $X \subseteq Y$ then polar$(X) \supseteq$ polar$(Y)$. Since $\mathcal{P}^{N}$ lies within the probability simplex, the polar of $\mathcal{P}^{N}$ contains the polar of the probability simplex, which is another simplex whose vertices are the distributions
\eqn{ \label{eqn:basis_states_again}
p_k(i) = (1-NL)\delta_{ik} + L \, , k \in \{1,\dots, N\} \, .
}
For a proof, see Lemma 4 in Ref.\ \cite{QPLEX}. And since $\mathcal{P}^{N}$ is self-polar, this set is contained in $\mathcal{P}^{N}$. Thus we have established that the state space contains the states of the form \eqref{eqn:basis_states_again}.

Next, we prove that {\bf A3} implies that these states have the maximum norm, as expressed by Eq.\  \eqref{eqn:extremality}. First note that the distributions \eqref{eqn:basis_states_again} exist on the surface of a sphere, since they have the same norm. We shall call this sphere and the vectors inside it the ``out-ball''. The out-ball is mutually polar with another ball, which happens to be the largest ball that can be inscribed inside the probability simplex (Lemma 6 in Ref.\ \cite{QPLEX}). By assumption {\bf A3}, this ball is the in-ball of $\mathcal{P}^{N}$, hence is fully contained within $\mathcal{P}^{N}$. Since polarity reverses inclusion, it follows that $\mathcal{P}^{N}$ is fully contained within the polar of the in-ball, that is, within the out-ball. Since the vectors $\{p_k \}$ are on the surface of this ball, they must have the maximum possible norm, that is $U$. $\Box$

This logic also works in reverse. Thanks to self-polarity, we can make an assumption either about the largest ball contained within the state space or about the smallest ball that contains it. Instead of adopting {\bf A3}, we could postulate a condition {\bf A3}${\bf '}$ declaring that the basis states lie on the sphere that just encloses the state space. This would fix $U$ in terms of $N$ and $L$. Again, this can be motivated by parsimony, since it means avoiding the introduction by hand of a new distance scale.

Along the way, we have also proven that the physically valid vectors that can become rows in measurement matrices are, up to scaling, the physically valid states. In other words, any vector in $\mathcal{M}^N$ is a prefactor times some vector in $\mathcal{P}^N$. This is the probabilistic statement of the fact that in quantum theory, any effect operator in a POVM becomes a density matrix when renormalized by its trace. Sometimes called ``self-duality'', this condition follows from the Born rule \eqref{eqn:matrix_urg} and the assumption {\bf A2} of maximality with respect to upper and lower bounds. In order to be physically valid, a vector $r_j$ must have a nonnegative inner product with $\Phi p$ for all $p \in \mathcal{P}^N$. Write
\eqn{
  r_j(i) = \alpha s_j(i)
}
where $\alpha > 0$ and $s_j$ is a properly normalized probability vector. Then the condition that the Born rule must give nonnegative values implies that
\eqn{
  \zum{i=1}{N} p(i) s_j(i) \geq L \, .
}
So, when a vector $r_j$ is normalized to have unit sum, the result must lie in the polar of $\mathcal{P}^N$, which is just $\mathcal{P}^N$.

\section{Consequences of {\bf A4} \label{app:vitality4}}
Within the set of possible states $\mathcal{P}^N$, recall that a ``mutually maximally distant (MMD) set of states'' is defined as a set satisfying
\eqn{
(p_i, p_i) &=& U, \nonumber \\
(p_i, p_j) &=& L,  \quad  i \neq j.
}
The uniform distribution is a vector $\iota := \{\frac{1}{N} : i=1,\dots,N \}$ that necessarily lies inside $\mathcal{P}^N$ (a consequence of {\bf A2}). It will be convenient to use co-ordinates in which this vector is the origin, by shifting $p \mapsto p-\iota := p'$. In these shifted co-ordinates the MMD set satisfies:
\eqn{
(p'_i, p'_i) &=& U-\frac{1}{N} \nonumber \\
&:=& U', \nonumber \\
(p'_i, p'_j) &=& L-\frac{1}{N} \nonumber \\
&:=& -L',  \qquad  i \neq j.
}
Note that
\eqn{
L \leq (\iota,\iota) \leq U \nonumber \\
\Rightarrow L \leq \frac{1}{N} \leq U \, ,
}
so the quantities $U'$, $L'$ defined above are both strictly positive. Next consider the vector $V$ defined as the sum of all $m$ vectors in the MMD set:
\eqn{
V := \zum{i}{m} p'_i
}
The norm of $V$ is:
\eqn{ \label{eqn:normV}
(V,V) = m\, U'+(m^2-m)\, (-L') \, ,
}
and since the norm is necessarily nonnegative,
\eqn{ \label{eqn:mmd_ineq}
0 \leq m\, U'+(m^2-m)\, (-L') \nonumber \\
\Rightarrow m \leq 1+\frac{U'}{L'} \, .
}
Note that this bound is tight, i.e.\ there is a possible choice of $\mathcal{P}^N$ for which it is realized. On the other hand, by definition, the maximum possible size of an MMD set is the system's dimension:
\eqn{ \label{eqn:mmd_ineq_d}
m \leq d \, .
}
Identifying these bounds leads to
\eqn{ \label{eqn:m_d_bound}
d &=& 1+\frac{U'}{L'} \, \nonumber \\
\Rightarrow U &=& (1-d)L+\frac{1}{N}d \, .
}
Furthermore, recall that the reference states have the form
\eqn{
e_k(i) = (1-NL)\delta_{ik} + L \, , k \in \{1,\dots, N\} \, ,
}
and have maximal norm, hence
\eqn{ \label{eqn:U_maxnorm}
U &=& (e_k, e_k) \nonumber \\
&=& (1-NL)^2 + 2L(1-NL) +NL^2 \nonumber \\
&=& 1-2NL+N^2L^2+2L-NL^2 \nonumber \\
&=& 1+L(N-1)(NL-2) \, .
}
By assumption {\bf A4} we have $N=d^2$; substituting this into Eqs.\ \eqref{eqn:m_d_bound}, \eqref{eqn:U_maxnorm} yields $L = 1/(d^2+d)$ and $U = 2L$. $\Box$

\section{Proof of the Lemma \label{app:prop2}}
A version of this result was first derived in Refs.\ \cite{BUSCH, RENES} to prove Gleason's theorem for POVMs. Our version applies to probability vectors instead of POVM elements, but is otherwise very similar.

Our strategy is to show that there exists an extension of $f$ to the full vector space $\mathbb{R}^N$, which satisfies the linearity property \eqref{eqn:prop2} on the whole space. Since the extension reduces to $f$ when restricted to the original domain $\mathcal{V}^N$, $f$ must be also be linear.

First we note that if $v$ is in $\mathcal{V}^N$, then so must be $\frac{1}{n} v$, for any positive integer $n$. (To see why, just consider that any outcome of an apparatus can be ``fine-grained'' by appending to it the outcome of the $n$-outcome ``garbage disposal'' apparatus.
Thus if an outcome originally occurred with probability $v$, it is fine-grained into a set of $n$ outcomes that each have probability $\frac{1}{n} v$). Then using the additivity of $f$ we have
\eqn{
f(v) = n f\left(\frac{1}{n}  v \right)
=  m f\left(\frac{1}{m}  v \right) \, ,
}
for arbitrary positive integers $n,m$. If we define $v':= \frac{1}{m} v$ we obtain
\eqn{
f\left(\frac{m}{n}  v' \right)
=  \frac{m}{n} f\left( v' \right) \,
}
and hence $f(a v) = a f(v)$ for any positive rational $a$.

Since $\mathcal{V}^N$ spans $\mathbb{R}^{+ N}$, any vector of positive components $w \in \mathbb{R}^{+ N}$ can be written as $w = a v$ for some $v \in \mathcal{V}^N$, where $a$ is positive and rational. Hence we can define an extension of $f$ to all positive vectors as: $f^+(w) := a f(v)$. Note that this definition is independent of the particular choice of  decomposition of $w$. For suppose that $w = a_1 v_1 = a_2 v_2$. Then $v_2 = \frac{a_1}{a_2} v_1$, so
\eqn{
f(v_2) &=& f\left( \frac{a_1}{a_2} v_1 \right) \nonumber \\
&=& \frac{a_1}{a_2} f\left( v_1 \right) \,
}
and therefore $f^{+}(w) := a_1 f(v_1) = a_2 f(v_2)$. Next we show that this extension is additive. Let $u,w$ be any vectors in $\mathbb{R}^{+ N}$. Since $\mathcal{V}^N$ spans $\mathbb{R}^{+ N}$, there exists some rational $a \geq 1$ such that the vectors $\frac{1}{a}(w+u),\frac{1}{a}w, \frac{1}{a}u$ are all in $\mathcal{V}^N$. Thus:
\eqn{
f^{+}(w+u) &=& a f^{+}\left( \frac{1}{a}(w+u) \right) \nonumber \\
&=& a f\left( \frac{1}{a}(w+u) \right) \nonumber \\
&=& a f\left( \frac{1}{a}w \right) + a f\left( \frac{1}{a}u \right) \nonumber \\
&=& a f^{+}\left( \frac{1}{a}w \right) + a f^{+}\left( \frac{1}{a}u \right) \nonumber \\
&=& f^{+}\left( w \right) +  f^{+}\left( u \right) \, .
}
To further extend our function to include vectors with negative components, note that any vector $z \in \mathbb{R}^{N}$ can be written as $z = w-u$ for positive $w,u \in \mathbb{R}^{+ N}$. Hence we may define the extension as $ \tilde{f}(z) := f^{+}(w)-f^{+}(u)$. Notice that by its very definition this extension is additive. To see that it does not depend on the choice of decomposition of $z$ into $w,u$, suppose that $z = w_1 - u_1 = w_2 - u_2$, hence $w_1+u_2 = w_2+u_1$. Then:
\eqn{
\tilde{f}(w_1+u_2) &=& \tilde{f}(w_2+u_1) \nonumber \\
\Rightarrow f^{+}(w_1+u_2) &=& f^{+}(w_2+u_1) \nonumber \\
\Rightarrow f^{+}(w_1)+f^{+}(u_2) &=& f^{+}(w_2)+f^{+}(u_1) \nonumber \\
\Rightarrow f^{+}(w_1)-f^{+}(u_1) &=& f^{+}(w_2)-f^{+}(u_2) \, .
}
Comparing the LHS and RHS we see that the extension must be the same, regardless of whether it is defined using the decomposition $z = w_1 - u_1$ or $z = w_2 - u_2$.

We have now shown that $f$ can be extended to an additive function $\tilde{f}$ on the full vector space $\mathbb{R}^{N}$. To prove that $\tilde{f}$ is linear, first note that:
\eqn{ \label{eqn:almost_linear}
\tilde{f}\left( \zum{x}{}\, \alpha_x v_x \right) &=& \zum{x}{}\, \tilde{f}\left( \alpha_x v_x \right) \, .
}
Moreover, since $\tilde{f}$ is additive, a similar argument as was used above for $f$ can be applied to prove that $\tilde{f}(\alpha_x v_x) = \alpha_x \tilde{f}(v_x)$ for any rational $\alpha_x$. What if $\alpha_x$ are irrational? Consider $x,z \in \mathbb{R}^N$ with $x \leq z$ (i.e., $y := z - x$ is entrywise nonnegative). Then $x + y = z$ and the additivity of $\tilde{f}$ gives $\tilde{f}(x) \leq \tilde{f}(z)$. Let $\alpha$ be an irrational number, and let $\{a_n : n = 1,2,\dots \}$ be an increasing sequence and $\{b_n : n = 1,2,\dots \}$ a decreasing sequence of rational numbers that both converge to $\alpha$. It follows that for any entrywise nonnegative $z$,
\eqn{
&& \tilde{f}(a_n z) \leq \tilde{f}(\alpha z) \leq \tilde{f}(b_n z) \nonumber \\
&\Rightarrow& a_n \tilde{f}(z) \leq \tilde{f}(\alpha z) \leq b_n \tilde{f}(z) \, .
}
Since $a_n \tilde{f}(\alpha z)$ and $b_n \tilde{f}(\alpha z)$ approach the same limit, by the ``pinching theorem'' of calculus this limit must be $\tilde{f}(\alpha z)$. Hence we can consistently define $\tilde{f}(\alpha z) := \alpha \tilde{f}(z)$ for any real $\alpha$. Applying this to the last line of \eqref{eqn:almost_linear}, we finally obtain that the function $\tilde{f}$ is linear:
\begin{equation}
\tilde{f}\left( \sum_x \, \alpha_x v_x \right) = \sum_x \, \alpha_x \tilde{f}\left( v_x \right) \, .
\end{equation}

\ \\

Finally, we conclude the proof by observing that by definition $\tilde{f} = f$ when restricted to the original domain $\mathcal{V}^N$, and so it follows that $f$ is linear on its domain, as desired. $\Box$

\bigskip

This proof somewhat parallels Wright and Weigert's proof of a Gleason-type theorem for ``General Probabilistic Theories''~\cite{WRIGHT}; both our theorem and theirs have the Busch and Renes et al.\ papers as common ancestors. Our program of probabilistic representations of quantum theory differs from the GPT tradition by, for example, proving the convexity of state space rather than assuming it, and requiring only one reference measurement for each $N$ rather than a family that must be considered conjointly. In essence, the GPT school abstracts the notion of von Neumann measurements, whereas we start with informationally complete POVMs.

\end{appendix}

\end{document}